\documentclass[10pt]{article}

\begin{document}
\input amssym.tex

\title{ISOMETRY GENERATORS IN MOMENTUM REPRESENTATION OF THE DIRAC THEORY ON THE DE SITTER SPACETIME}

\author{ION I. COTAESCU \footnote{e-mail: icotaescu@yahoo.com}\\DORU-MARCEL BALTATEANU \footnote{e-mail: bdmarcel@yahoo.com }\\{\it West University of Timi\c soara, V. Parvan Ave. 4,}\\
{\it Timi\c soara, RO-300223, Romania }}

\maketitle


\begin{abstract}
In this paper, it is shown that the covariant representation (CR) transforming the Dirac field under de Sitter isometries is equivalent to a direct sum of two unitary irreducible representations (UIRs) of the $Sp(2,2)$ group transforming alike the particle and antiparticle field operators in momentum representation. Their basis generators and Casimir operators are written down for the first time finding that these representations are equivalent  to  a UIR from the principal series whose canonical  labels are determined by the fermion mass and spin. The properties of the conserved observables (i. e. one-particle operators) associated to the de Sitter isometries via Noether theorem and of the corresponding  Pauli-Lubanski type operator are also pointed out.

Pacs: 04.62.+v
\end{abstract}


\section{Introduction}

In  general relativity there are covariant quantum fields defined on a curved manifold $M$ which transform under isometries according to CRs  {\em induced} by finite representations of the universal covering group $\hat G$ of the gauge one, $G$ \cite{ES,ES1}. In the case of  four-dimensional local-Minkowskian manifolds under consideration here, these groups are $G=SO(1,3)$ and respectively $\hat G=SL(2,{\Bbb C})$.  For this reason, the spin terms of the operators generating CRs are given by the linear representations of the $sl(2,{\Bbb C})$ algebra instead of the  $s(M)$ algebra of the universal covering group $S(M)$ of the isometry one, $I(M)$.  

On the other hand, the group $S(M)$ has UIRs that  may be related to the CRs. In special relativity, the CRs are equivalent to orthogonal sums of  Wigner's UIRs  that govern the transformation rules under isometries of the particle and antiparticle operators in momentum representation \cite{WKT}. This result seems to be related to the special structure of the  Poincar\' e  isometry group, $G\circledS T(4)$ such that it cannot be generalized to other manifolds, not even in the case of the de Sitter spacetime  which still allows a momentum representation.  

The de Sitter manifold, denoted from now by $M$, is local-Minkowskian and has the isometry group $I(M)=SO(1,4)$ which is the gauge group of the Minkowskian five-dimensional manifold $M^5$ embedding $M$. The UIRs of the corresponding group  $S(M)={\rm Spin}(1,4)=Sp(2,2)$ are well-studied \cite{linrep,linrep1} and used in various applications. Many authors exploited this high symmetry for building quantum theories,  either by constructing symmetric two-point functions, avoiding thus the canonical quantization \cite{Wood,Wood1}, or by using directly these UIRs for finding field equations without considering CRs \cite{Gaz,Gaz1,Gaz2}. Another approach which applies the {\em canonical quantization} to the covariant fields transforming according to {\em induced} CRs was initiated by Nachtmann \cite{Nach}  many years ago and continued in few of our papers \cite{CNach,CPF,CSB,CCC,CQED}. 

Here we adopt this last framework for investigating the relation among the CRs and UIRs of the Dirac theory in momentum representation on the de Sitter background.  As mentioned,  these CRs  are induced by the linear representations of the group $\hat G$ without to meet explicitly  the linear representations of the group $S(M)$.  Nevertheless, by studying the generators of  the CRs and the corresponding Casimir operators we found indirectly that these CRs may be equivalent to orthogonal sums of UIRs of the group $S(M)$,  but without writing down the generators of the UIRs in momentum representation \cite{CCC}.  Now  we would like to continue this study showing that the CRs-UIRs equivalence can be demonstrated at least in the case of the free Dirac field on $M$ for which we succeed to construct the UIRs in momentum representation deriving their basis generators.  With their help we obtain  the conserved one-particle operators associated to the de Sitter isometries that are the principal observables of the quantum field theory (QFT).

We specify that the generators of the scalar UIR in this representation were deduced by Nachtmann \cite{Nach}.  Therefore,  what is new here are the spin terms of the generators of spinor UIRs and the Casimir operators that can be calculated resorting to suitable algebraic codes on computer. In this manner we find that the Dirac particle and antiparticle operators in momentum representation transform according to the {\em same} UIR that can be one of the equivalent UIRs $(s,q)$,  from the principal series \cite{linrep,linrep1},  labeled by  the spin $s=\frac{1}{2}$ and $q=\frac{1}{2}\pm  i\frac{m}{\omega}$ where $m$ is the fermion mass and $\omega$ is the  Hubble constant of $M$  in our notation. 

This result is new and similar to that of special relativity where the particle and antiparticle operators in momentum representation of any covariant quantum field with unique spin, $s$, transform alike under Poincar\'e isometries, according to the same Wigner UIR induced by the $2s+1$-dimensional UIR  of the group $SU(2)$ \cite{W}. Obviously, this happens only if we respect the spin-statistic connection, assuming that  the Dirac particle and antiparticle operators satisfy canonical anti-commutation rules. 

On the other hand, the concrete forms of the generators derived here for the first time allow us to expand  in momentum representation  the conserved one-particle operators corresponding to the de Sitter isometries via Noether theorem. We show that for all these operators (energy, momentum, angular momentum, etc.) the contributions of the particles and antiparticles are {\em additive} in contrast with the conserved charge where these have opposite signs.  Thus we hope that the new results presented here should complete our theory of the quantum Dirac  field on $M$ \cite{CPF,CSB,CCC}, giving all the technical details needed in various applications in QFT.

Note that the fundamental spinors we use here correspond to a fixed vacuum of the Bounch-Davies type \cite{BD}  as in our de Sitter QED \cite{CQED}. They are defined up to some arbitrary phase factors which could induce supplemental terms in generator's expressions.  

This paper is organized as follows. After a short presentation,  in the next section, of the fundamental solutions of the free Dirac equation on $M$, in the third one we derive the generators of the UIRs in momentum representation and the components of the Pauli-Lubanski operator which helps us to obtain the Casimir operators.  Moreover, we give the general form of the momentum expansions of the principal conserved observables of  QFT discussing briefly their properties. Finally, we present our conclusions.

\section{Dirac field on the de Sitter spacetime}

In what follows we focus on the Dirac field on the  expanding portion of $M$  where we consider  the frames $\{t,\vec{x};e\}$ formed by the conformal Euclidean chart $\{t,\vec{x}\}$, with the conformal flat line element \cite{BD},
$ds^{2}=(\omega^2 {t})^{-2}\left({dt}^{2}-d\vec{x}^2\right)$,
in the domain  $t\in (-\infty,\,0)$  and $\vec{x}\in D= {\Bbb R}^3$, and 
the local  frames defined by the diagonal gauge, $e^{0}_{0}=-\omega t$,  $e^{i}_{j}=-\delta^{i}_{j}\,\omega t$ and $\hat e^{0}_{0}=-(\omega t)^{-1}$, $\hat e^{i}_{j}=-\delta^{i}_{j}
(\omega t)^{-1}$ ($i,j,...=1,2,3$). In this frame, the free Dirac equation \cite{CPF},
\begin{equation}\label{EDir}
(E_D-m)\psi(x)=\left[-i\omega t\left(\gamma^0\partial_{t}+\gamma^i\partial_i\right)
+\frac{3i\omega}{2}\gamma^{0}-m\right]\psi(x)=0\,,
\end{equation}
depends on the point-independent Dirac matrices $\gamma^{\hat\mu}$ that satisfy $\{ \gamma^{\hat\alpha},\, \gamma^{\hat\beta} \}=
2\eta^{\hat\alpha \hat\beta}$ giving rise to the basis-generators
$S^{\hat\alpha \hat\beta}=\frac{i}{4} [\gamma^{\hat\alpha}, \gamma^{\hat\beta}
]$ of the spinor representation $(\frac{1}{2},0)\otimes (0,\frac{1}{2})$ of the group $\hat G=SL(2,\Bbb C)$ that induces the spinor CR \cite{ES,CPF,CSB}. 

Eq. (\ref{EDir}) can be analytically solved either in momentum or energy bases with correct orthonormalization and completeness properties \cite{CPF,CSB} with respect to the relativistic scalar product 
\begin{equation}
\left< \psi,\psi'\right>=\int d^3x\,(-\omega t)^{-3}\overline{\psi}(t,\vec{x})\gamma^0 \psi'(t,\vec{x})\,.
\end{equation}
The mode expansion in the spin-momentum representation \cite{CSB},
\begin{equation}\label{psiab}
\psi(t,\vec{x})=\int d^3 p \sum_{\sigma}\left[U_{\vec{p},\sigma}(x)
a(\vec{p},\sigma)+V_{\vec{p}, \sigma}(x){b}^{\dagger}(\vec{p},
\sigma) \right]\,,
\end{equation}
is written in terms of the field operators, $a$ and $b$ (satisfying canonical anti-commutation rules), and  the particle and antiparticle fundamental spinors of momentum $\vec{p}$ (with $p=|\vec{p}|$) and polarization $\sigma=\pm\frac{1}{2}$, 
\begin{equation}
U_{\vec{p},\sigma}(t,\vec{x}\,)=\frac{1}{(2\pi)^{\frac{3}{2}}}\, u_{\vec{p},\sigma}(t) e^{i\vec{p}\cdot\vec{x}}\,,\quad
V_{\vec{p},\sigma}(t,\vec{x}\,)=\frac{1}{(2\pi)^{\frac{3}{2}}}\, v_{\vec{p},\sigma}(t) e^{-i\vec{p}\cdot\vec{x}}\,,
\end{equation}
whose time-dependent terms have the form \cite{CSB,CQED}
\begin{eqnarray}
u_{\vec{p},\sigma}(t)&=&  \frac{i}{2}\left(\frac{\pi p}{\omega}\right)^{\frac{1}{2}}(\omega t)^2\left(
\begin{array}{r}
e^{\frac{1}{2}\pi\mu}H^{(1)}_{\nu_{-}}(-p t) \,
\xi_{\sigma}\\
e^{-\frac{1}{2}\pi\mu}H^{(1)}_{\nu_{+}}(-p t) \,\frac{\vec{\sigma}\cdot\vec{p}}{p}\,\xi_{\sigma}
\end{array}\right)\,,
\label{Ups}\\
v_{\vec{p},\sigma}(t)&=&\frac{i}{2} \left(\frac{\pi p}{\omega}\right)^{\frac{1}{2}}(\omega t)^2 \left(
\begin{array}{r}
e^{-\frac{1}{2}\pi\mu}H^{(2)}_{\nu_{-}}(-p t)\,\frac{\vec{\sigma}\cdot\vec{p}}{p}\,
\eta_{\sigma}\\
e^{\frac{1}{2}\pi\mu}H^{(2)}_{\nu_{+}}(-p t) \,\eta_{\sigma}
\end{array}\right)
\,,\label{Vps}
\end{eqnarray}
in the standard representation of the Dirac matrices (with diagonal $\gamma^0$) and a fixed vacuum of the Bounch-Davies type \cite{CQED}. Obviously, the notation $\sigma_i$ stands for the Pauli matrices while  the point-independent Pauli spinors $\xi_{\sigma}$ and $\eta_{\sigma}= i\sigma_2 (\xi_{\sigma})^{*}$ are  normalized as $\xi^+_{\sigma}\xi_{\sigma'}=\eta^+_{\sigma}\eta_{\sigma'}=\delta_{\sigma\sigma'}$ \cite{CSB}. The terms giving the time modulation depend on the Hankel functions $H_{\nu_{\pm}}^{(1,2)}$   of indices 
\begin{equation}
\nu_{\pm}=\frac{1}{2}\pm i\mu \,, \quad  \mu=\frac{m}{\omega}\,.  
\end{equation}
Based on their properties (presented in the Appendix A) we deduce
\begin{equation}
{u}^+_{\vec{p},\sigma}(t){u}_{\vec{p},\sigma}(t)={v}^+_{\vec{p},\sigma}(t){v}_{\vec{p},\sigma}(t)=(-\omega t)^3
\end{equation}
obtaining  the ortonormalization relations \cite{CPF} 
\begin{eqnarray}
&&\left<U_{\vec{p},\sigma},U_{\vec{p}^{\,\prime},\sigma^{\prime}}\right>=
\left<V_{\vec{p},\sigma},V_{\vec{p}^{\,\prime},\sigma^{\prime}}\right>=
\delta_{\sigma\sigma^{\prime}}\delta^3 (\vec{p}-\vec{p}^{\,\prime})\,,
\label{orto1}\\
&&\left<U_{\vec{p},\sigma},V_{\vec{p}^{\,\prime},\sigma^{\prime}}\right>=
\left<V_{\vec{p},\sigma},U_{\vec{p}^{\,\prime},\sigma^{\prime}}\right>=
0\,,\label{orto2}
\end{eqnarray}
that yield  the useful  inversion formulas, 
$a(\vec{p},\sigma)=\left<U_{\vec{p},\sigma},\psi\right>$ and 
$b(\vec{p},\sigma)=\left<\psi,V_{\vec{p},\sigma}\right>$. Moreover, it is not hard to verify that these spinors are charge-conjugated to each other,
\begin{equation}\label{conj}
V_{\vec{p},\sigma}=(U_{\vec{p},\sigma})^{c}={\cal C}
(\overline{U}_{\vec{p},\sigma})^T \,, \quad {\cal C}=i\gamma^2\gamma^0\,,
\end{equation}
and represent a {\em complete} system of solutions in the sense that \cite{CPF}
\begin{equation}\label{compl}
\int d^3 p \sum_{\sigma}\left[
U_{\vec{p},\sigma}(t,\vec{x})U^{+}_{\vec{p},\sigma}(t,\vec{x}^{\,\prime})+
V_{\vec{p},\sigma}(t,\vec{x})V^{+}_{\vec{p},\sigma}(t,\vec{x}^{\,\prime})
\right]=e^{-3\omega t}\delta^3 (\vec{x}-\vec{x}^{\,\prime})\,.
\end{equation}

The Dirac field transforms under isometries $x\to x'=\phi_{\frak g}(x)$ (with ${\frak g}\in I(M)$) according to the CR  $T_{\frak g} : \psi(x) \to (T_{\frak g}\psi)(x')=A_{\frak g}(x)\psi(x)$ where 
$A_{\frak g}(x)\in SL(2,{\Bbb C})$ is defined as in Ref. \cite{ES}. The above inversion formulas allow us to write the transformation rules in momentum representation as
\begin{equation}\label{Tg}
(T_{\frak g}a)(\vec{p},\sigma)=\left<U_{\vec{p},\sigma},(A_{\frak g}\psi)\circ \phi^{-1}_{\frak g}\right>\,, \quad (T_{\frak g}b)(\vec{p},\sigma)=\left<(A_{\frak g}\psi)\circ \phi^{-1}_{\frak g},V_{\vec{p},\sigma}\right>\,,
\end{equation}
but, unfortunately, these scalar product are complicated integrals that cannot be solved.  Therefore, we must restrict ourselves to study the corresponding Lie algebras focusing on the basis generators of these representations. 

\section{Isometry generators  in momentum representation} 

In the covariant parametrization of the $sp(2,2)$ algebra  the generators $X_{(AB)}$ corresponding to the Killing vectors $k_{(AB)}$ ($A,B,...=0,1,2,3,4$) can be calculated as in Refs. \cite{CPF,CCC}.  These operators are the energy (or Hamiltonian) $H$,  total angular momentum $\vec{J}$, generators of the Lorentz boosts $\vec{K}$, and  Runge-Lenz type vector $\vec{R}$, whose components read \cite{CCC},
\begin{eqnarray}
H& =&  \omega X_{(04)}=-i\omega(t \partial_t + {x}^i
{\partial}_i)\,,\label{Ham}\\
J_i &=&  \frac{1}{2}\,\varepsilon_{ijk}X_{(jk)}=-i\varepsilon_{ijk}x^j\partial_k+S_i\,,\quad S_i=\frac{1}{2}\varepsilon_{ijk}S_{jk}\,,\label{Ji}\\
K_i& =&  X_{(0i)}=i x^i H+
\frac{i}{2\omega}[1+\omega^2( \vec{x}^2-t^2)]\partial_i -\omega
tS_{0i}+\omega S_{ij}x^j\,,\label{Ki}\\
R_i  &=&X_{(i4)}=-K_i+\frac{1}{\omega}\, i\partial_i\,.
\end{eqnarray}
The momentum operator, $\vec{P}$,  and its dual, $\vec{Q}$, having the components
\begin{equation}\label{Pi}
P_i=- \omega(R_i+K_i)=-i\partial_i\,, \quad
Q_i= \omega(R_i-K_i)\,,
\end{equation}
form two Abelian subalgebras of the $sp(2,2)$ algebra.

Any self-adjoint generator $X$ of the spinor representation of the group $S(M)$  gives rise to a {\em conserved} one-particle operator of the QFT, 
\begin{equation}\label{X}
{\bf X}=:\left<\psi, X\psi\right>:={\bf X}^{(+)}+{\bf X}^{(-)}=
\int d^3 p\, 
\left[\alpha^{\dagger}(\vec{p}) {\tilde X}^{(+)}\alpha(\vec{p})
+\beta^{\dagger}(\vec{p})\tilde X^{(-)} \beta(\vec{p})\right]\,,
\end{equation}
calculated respecting the normal ordering of the operator products \cite{BDR}.  
The operators ${\tilde X}^{(\pm)}$ are the generators of CRs in momentum representation acting on the operator valued Pauli spinors, 
\begin{equation}
\alpha(\vec{p})=\left(\begin{array}{l}
a(\vec{p},\frac{1}{2})\\
a(\vec{p},-\frac{1}{2})
\end{array}\right)\,,\quad
\beta(\vec{p})=\left(\begin{array}{l}
b(\vec{p},\frac{1}{2})\\
b(\vec{p},-\frac{1}{2})
\end{array}\right) \,.
\end{equation}
As observed in Ref. \cite{Nach}, the straightforward method for finding the structure of these operators is to evaluate the entire expression  (\ref{X}) by using the form  (\ref{psiab}) where the field operators $a$ and $b$ satisfy the canonical antcommutation rules \cite{Nach,CPF}.

For this purpose we consider several identities written with the notation $\partial_{p_i}=\frac{\partial}{\partial p_i}$ as 
\begin{eqnarray}
&x^i\,U_{\vec{p},\sigma}(t,\vec{x})=-i\partial_{p_i}   U_{\vec{p},\sigma}(t,\vec{x}) \,, & H\,U_{\vec{p},\sigma}(t,\vec{x})=-i\omega
\left(p^i\partial_{p^{i}}+\frac{3}{2}\right)
U_{\vec{p},\sigma}(t,\vec{x})\,,\nonumber\\
&x^i\,V_{\vec{p},\sigma}(t,\vec{x})=i\partial_{p_i}   V_{\vec{p},\sigma}(t,\vec{x})         \,,& H\,V_{\vec{p},\sigma}(t,\vec{x})=-i\omega
\left(p^i\partial_{p^{i}}+\frac{3}{2}\right)
V_{\vec{p},\sigma}(t,\vec{x})\,,\nonumber
\end{eqnarray}
that help us to eliminate some multiplicative operators and the time derivative when we  inverse the Fourier transform. Furthermore, by applying the Green theorem and calculating on computer terms of the form ${u}^+_{\vec{p},\sigma}(t)F(t,{p}_i){u}_{\vec{p},\sigma}(t)$,  ${v}^+_{\vec{p},\sigma}(t)F(t,{p}_i){v}_{\vec{p},\sigma}(t)$,..., etc., we find two  {\em identical} representations whose basis generators read,   ${\tilde P}^{(\pm)}_i=\tilde P_i=p_i$ and
\begin{eqnarray}
{\tilde H}^{(\pm)}&=&\omega {\tilde X}^{(\pm)}_{(04)}=i \omega \left (p_i \partial_{p_i}+\frac{3}{2}\right)\label{Hamp}\,,\\
{\tilde J_i}^{(\pm)}&=& \frac{1}{2}\,\varepsilon_{ijk}
{\tilde X}^{(\pm)}_{(jk)}=-i\varepsilon_{ijk}p_j\partial_{p_k}+\frac{1}{2}\sigma_i\,,\\
{\tilde K}^{(\pm)}_i&=&{\tilde X}^{(\pm)}_{(0i)}= i {\tilde H}^{(\pm)}\partial_{p_i}+\frac{\omega}{2}p_i\Delta_p - p_i \frac{\vec{p}^2+m^2}{2\omega \vec{p}^2}\nonumber\\
&&~~~~~~~~~~+\frac{1}{2}\varepsilon_{ijk}\left(i\omega \partial_{p_j}- p_j\frac{m}{2\vec{p}^2}\right)\sigma_k\,,\\
{\tilde R}^{(\pm)}_i&=&{\tilde X}^{(\pm)}_{(i4)}=-{\tilde K}^{(\pm)}_i-\tilde P_i\,,\label{Rip}
\end{eqnarray}
where $\Delta_p=\partial_{p_i}\partial_{p_i}$.
These basis generators satisfy the specific  $sp(2,2)$ commutation rules:
\begin{eqnarray} &\left[{\tilde J}^{(\pm)}_i,
{\tilde J}^{(\pm)}_j\right]=i\varepsilon_{ijk}{\tilde J}^{(\pm)}_k\,,\quad&\left[{\tilde J}^{(\pm)}_i,
{\tilde R}^{(\pm)}_j\right]=i\varepsilon_{ijk}{\tilde R}^{(\pm)}_k\,,\label{ALG1}\\
&\left[{\tilde J}^{(\pm)}_i, {\tilde K}^{(\pm)}_j\right]=i\varepsilon_{ijk}
{\tilde K}^{(\pm)}_k\,,\quad&\left[{\tilde R}^{(\pm)}_i,
{\tilde R}^{(\pm)}_j\right]=i\varepsilon_{ijk}
{\tilde J}^{(\pm)}_k\,,\label{ALG2}\\
&\left[{\tilde K}^{(\pm)}_i,{\tilde  K}^{(\pm)}_j\right]=-i\varepsilon_{ijk}
{\tilde J}^{(\pm)}_k\,,\quad&\left[{\tilde R}^{(\pm)}_i,
{\tilde K}^{(\pm)}_j\right]=\frac{i}{\omega}\,\delta_{ij}{\tilde H}^{(\pm)}\,,\label{ALG3}
\end{eqnarray}
and
\begin{equation}\label{HHKR}
\left[{\tilde H}^{(\pm)}, {\tilde J}^{(\pm)}_i\right]=0\,,\quad \left[{\tilde H}^{(\pm)},
{\tilde K}^{(\pm)}_i\right]=i\omega {\tilde R}^{(\pm)}_i\,,\quad\left[\tilde H^{(\pm)},
{\tilde R}^{(\pm)}_i\right]=i\omega{\tilde K}^{(\pm)}_i\,.
\end{equation}
Moreover, it is not difficult to verify that these are Hermitian operators with respect to the scalar products of the momentum representation
\begin{equation}
\left<\alpha, \alpha'\right>=
\int d^3 p\, 
\alpha^{\dagger}(\vec{p}) \tilde \alpha(\vec{p})\,, \quad 
\langle \beta,\beta'\rangle=\int  d^3p \beta^{\dagger}(\vec{p})\tilde \beta(\vec{p})\,.
\end{equation}
Therefore, we can conclude that these operators generate a pair of  {\em unitary} representations of the  group $S(M)$. 

Since all the UIRs of the group $S(M)$ are classified \cite{linrep},  we can study  the equivalence and reducibility of these representations simply  by calculating the Casimir operators. The first  Casimir operator is quadratic,
\begin{equation}
\tilde{\cal C}_1=-\,
\frac{1}{2}\,\tilde X_{(AB)}\tilde X^{(AB)}\label{Q1}\,,
\end{equation}
while the second one,
\begin{equation}\label{Q2}
\tilde{\cal C}_2=-\eta^5_{AB}\tilde W^{A}\tilde W^{B}\,,\quad \tilde W^{A}=\frac{1}{8}\, \varepsilon^{ABCDE}
\tilde X_{(BC)}\tilde X_{(DE)}\,,
\end{equation}
is written in terms of the Pauli-Lubanski operator of components $\tilde W^A$ by using the total anti-symmetric tensor with $\varepsilon^{01234}=1$ and the metric tensor $\eta^5={\rm diag}(1,-1,-1,-1,-1)$  of $M^5$.  After performing the calculation on computer we find
\begin{eqnarray}
\tilde W_0^{(\pm)}&=&\frac{\omega}{4}(\vec{\sigma}\cdot\vec{p})\Delta_p+\frac{\omega\nu_{-}}{2}\vec{\sigma}\cdot \vec{\partial}_p+\frac{i m}{2\vec{p}^2}(\vec{\sigma}\cdot \vec{p})\, \vec{p}\cdot\vec{\partial}_p\nonumber\\
&&+\frac{m^2-\vec{ p}^2+ 2i\omega m}{4 \vec{p}^2\omega}\vec{\sigma}\cdot\vec{p}\,,\label{W0}\\
\tilde W_i^{(\pm)}&=&\frac{i}{2}(\vec{\sigma}\cdot\vec{p}) \partial_{p_i}-\frac{i\nu_{-}}{2\vec{p}^2}\sigma_i - \frac{m}{2\omega\vec{p}^2}(\vec{\sigma}\cdot\vec{p}) p_i\,, \\
\tilde W_4^{(\pm)}&=&\tilde W_0^{(\pm)}+\frac{1}{2 \omega}\vec{\sigma}\cdot\vec{p}\,.\label{W4}
\end{eqnarray}
With these preparation we obtain the Casimir operators (\ref{Q1}) and (\ref{Q2}) as 
\begin{eqnarray}
\tilde{\cal C}^{(\pm)}_1&=&-s(s+1)-(q+1)(q-2)=  \mu^2+\frac{3}{2}\,, \\
\tilde {\cal C}^{(\pm)}_2&=&-s(s+1)q(q-1)=
s(s+1)\nu_+\nu_-=
\frac{3}{4}\left(\mu^2+\frac{1}{4}\right)\,,
\end{eqnarray}
which show  that the above identical spinor representations are  UIRs of the principal series corresponding to the canonical labels $(s,q)$ with  $s=\frac{1}{2}$ and $q=\nu_{\pm}$.  This suggests that the UIRs  $(s, \nu_{\pm})$ of the group $S(M)=Sp(2,2)$ can be seen as being analogous to the Wigner ones $(s,\pm m)$  of the Dirac theory in Minkowski spacetime. Note that a similar result was obtained in Ref. \cite{CCC} indirectly stating with the generators (\ref{Ham})-(\ref{Ki}) and exploiting the relation among the Casimir operators and the Dirac one in the configuration space.

In general, the above spinor representations may not coincide since  the expressions of their basis generators are strongly dependent on the arbitrary phase factors of the fundamental spinors whether these depend on $\vec{p}$. Thus if we change   
\begin{equation}\label{gaugeUV}
U_{\vec{p},\sigma}\to e^{i\chi^+(\vec{p})} U_{\vec{p},\sigma}\,,\quad
V_{\vec{p},\sigma}\to e^{-i\chi^-(\vec{p})} V_{\vec{p},\sigma}\,,
\end{equation}
with $\chi^{\pm}(\vec{p})\in \Bbb R$, performing simultaneously  the associated transformations,
\begin{equation}\label{gaugeab}
\alpha(\vec{p})\to e^{-i\chi^+(\vec{p})}\alpha(\vec{p})\,,\quad
\beta(\vec{p})\to e^{-i\chi^-(\vec{p})}\beta(\vec{p})\,,
\end{equation}
that preserves the form of $\psi$, we find that the operators ${\tilde P}_i$   keep their forms while the other generators are changing, e. g.  the Hamiltonian operators transform as ${\tilde H}^{(\pm)}\to  {\tilde H}^{(\pm)} + p^i \partial_{p^i}\chi^{\pm}(\vec{p})$. Obviously,  these transformations are nothing other than unitary transformations among equivalent UIRs.  Note that thanks to this mechanism one can fix  suitable phases for determining desired forms of the basis generators keeping thus under control the flat and rest limits of these operators in the Dirac \cite{CSB} or scalar \cite{Nach,Cscalar} field theory on $M$.

At the level of  QFT, the operators $\{{\bf X}_{(AB)}\}$, given by Eq. (\ref{X}) where we introduce the differential operators (\ref{Hamp}) -(\ref{Rip}), generate a reducible operator valued CR which can be decomposed as the orthogonal sum of  CRs - generated by $\{{\bf X}_{(AB)}^{(+)}\}$ and $\{{\bf X}_{(AB)}^{(-)}\}$ - that are equivalent between themselves and equivalent to the UIRs $(\frac{1}{2},\nu_{\pm})$ of the $sp(2,2)$ algebra. These one-particle operators are the principal conserved quantities of the Dirac theory corresponding to the de Sitter isometries via Noether theorem. It is remarkable that in our formalism we have $\tilde X^{(+)}_{AB}=\tilde X^{(-)}_{AB}$ which means that the  particle and antiparticle sectors bring similar contributions such that we can say that these quantities are {\em additive}, e. g., the energy of a many particle system is the sum of the individual energies of particles and antiparticles.  

Other important conserved one-particle operators are the components of the Pauli -Lubanski operator,
\begin{equation}
{\bf W}_A={\bf W}^{(+)}_A+{\bf W}^{(-)}_A=
\int d^3 p\, 
\left[\alpha^{\dagger}(\vec{p}) {\tilde W}^{(+)}_A\alpha(\vec{p})
+\beta^{\dagger}(\vec{p})\tilde W^{(-)}_A \beta(\vec{p})\right]\,,
\end{equation}
as given by Eqs. (\ref{W0})-(\ref{W4}). The Casimir operators of QFT have to be calculated according to Eqs. (\ref{Q1}) and (\ref{Q2}) but by using the one-particle operators ${\bf X}_{(AB)}$ and ${\bf W}_A$ instead of ${\tilde X}_{(AB)}$ and ${\tilde W}_A$. We obtain
the following one-particle contributions
\begin{equation}
{\bf C}_1=\left (\mu^2+\frac{3}{2}\right){\bf N}+\cdots,\quad {\bf C}_2= \frac{3}{4}\left(\mu^2+\frac{1}{4}\right){\bf N}+\cdots,
\end{equation}
where ${\bf N}={\bf N}^{(+)}+{\bf N}^{(-)}$ is the usual operator of the total number of particles and antiparticles. Thus the additivity holds for the entire theory of the spacetime symmetries in contrast with the conserved charges of the internal symmetries that take different values for particles and antiparticles as, for example, the charge operator  corresponding to the $U(1)_{em}$ gauge symmetry \cite{CQED} that reads ${\bf Q}=q({\bf N}^{(+)}-{\bf N}^{(-)})$.

\section{Concluding remarks}

The first conclusion is that on the de Sitter background the spinor CR induced by the linear representation $(\frac{1}{2},0)\otimes (0,\frac{1}{2})$ of the group $\hat G=SL(2,\Bbb C)$ is equivalent to the orthogonal sum of two equivalent UIRs of the group $S(M)=Sp(2,2)$ labelled by $(\frac{1}{2},\nu_{\pm})$. Thus at least in the case of the Dirac field we recover a similar conjuncture as in  the Wigner theory of the induced representations of the Poincar\' e group in special relativity. However, the principal difference is that the transformations of  the Wigner UIRs  can be written in closed forms while in our case this cannot be done because of the technical difficulties  in solving the integrals (\ref{Tg}). For this reason we were forced to restrict ourselves to study only the representations  of the corresponding algebras.

 This is not an impediment since physically speaking we are interested  to know the properties of the basis generators (in configurations or momentum representation) since these give rise to the conserved observables (i. e. the one-particle operators) of QFT, associated to the de Sitter isometries. It is remarkable that the particle and antiparticle sectors of these operators bring additive contributions since  the particle and antiparticle operators transform alike under isometries  just as in special relativity. Notice that this result was obtained by Nachtmann \cite{Nach} for the scalar UIRs but this is less relevant as long as the generators of the scalar representation depend only on $m^2$. Now we see that  the generators of the spinor representation which have spin terms depending on $m$ preserve this property such that we can conclude that all the one-particle operators corresponding to the de Sitter isometries are additive,  regardless the spin.

In other respects, we observe that the fermion mass is defined by the Dirac equation such that we can express all our invariants  in terms of spin and mass in a similar manner as in special relativity. The challenge which remains is to generalize the above results to CRs of any spin but without using field equations as in Ref \cite{CCC} or their solutions as in the present paper.  This presumes to look for a general definition of  mass or even of a mass operator on $M$ related to the Casimir operators of the UIRs of the $S(M)$ group.  We note that despite of the well-known classical results \cite{Nach,BDurr} and some new interesting progresses \cite{Gmass} there remains a discrepancy between the manners in which the masses of bosons and fermions depend on the de Sitter invariants \cite{CCC}. We hope that the results presented here will offer one new tools in solving this delicate problem.

\appendix

\section{Some properties of Hankel functions}

According to the general properties of the Hankel functions \cite{GR}, we
deduce that those used here, $H^{(1,2)}_{\nu_{\pm}}(z)$, with
$\nu_{\pm}=\frac{1}{2}\pm i \mu$ and $z\in \Bbb R$, are related among
themselves through $[H^{(1,2)}_{\nu_{\pm}}(z)]^{*}
=H^{(2,1)}_{\nu_{\mp}}(z)$ and satisfy the  identities
\begin{equation}\label{H3}
e^{\pm \pi k}
H^{(1)}_{\nu_{\mp}}(z)H^{(2)}_{\nu_{\pm}}(z)
+ e^{\mp \pi k} H^{(1)}_{\nu_{\pm}}(z)H^{(2)}_{\nu_{\mp}}(z)=\frac{4}{\pi z}\,.
\end{equation}

\section*{Acknowledgments}

I.I. Cot\u aescu is supported by a grant of the Romanian National Authority for Scientific Research,
Programme for research-Space Technology and Advanced Research-STAR, project nr. 72/29.11.2013 between Romanian Space Agency and West University of Timisoara.

D.-M. B\u alt\u a\c teanu is supported by the strategic grant POSDRU /159/1.5/S /137750, Project “Doctoral and Postdoctoral programs support for increased competitiveness in Exact Sciences research”, cofinanced by the European Social Fund within the Sectoral Operational Programme Human Resources Development 2007 - 2013.

\end{document}